# Clues in Tweets: Twitter-Guided Discovery and Analysis of SMS Spam

Siyuan Tang
Indiana University Bloomington
tangsi@iu.edu

Xianghang Mi*
University of Science and Technology of China
xmi@ustc.edu.cn

Ying Li
SKLOIS, Institute of Information Engineering, CAS
liying1998@iie.ac.cn

XiaoFeng Wang
Indiana University Bloomington
xw7@indiana.edu

Kai Chen[†][‡]
SKLOIS, Institute of Information Engineering, CAS
chenkai@iie.ac.cn

## ABSTRACT

With its critical role in business and service delivery through mobile devices, SMS (Short Message Service) has long been abused for spamming, which is still on the rise today possibly due to the emergence of A2P bulk messaging. The effort to control SMS spam has been hampered by the lack of up-to-date information about illicit activities. In our research, we proposed a novel solution to collect recent SMS spam data, at a large scale, from Twitter, where users voluntarily report the spam messages they receive. For this purpose, we designed and implemented *SpamHunter*, an automated pipeline to discover SMS spam reporting tweets and extract message content from the attached screenshots. Leveraging *SpamHunter*, we collected from Twitter a dataset of 21,918 SMS spam messages in 75 languages, spanning over four years. To our best knowledge, this is the largest SMS spam dataset ever made public. More importantly, *SpamHunter* enables us to continuously monitor emerging SMS spam messages, which facilitates the ongoing effort to mitigate SMS spamming. We also performed an in-depth measurement study that sheds light on the new trends in the spammer's strategies, infrastructure and spam campaigns. We also utilized our spam SMS data to evaluate the robustness of the spam countermeasures put in place by the SMS ecosystem, including anti-spam services, bulk SMS services, and text messaging apps. Our evaluation shows that such protection cannot effectively handle those spam samples: either introducing significant false positives or missing a large number of newly reported spam messages.

## 1 INTRODUCTION

Although it has long been known that Short Message Service (SMS) is abused by cybercriminals for spamming, e.g., unwanted SMS advertising, SMS phishing (called *smishing*), etc., the problem becomes increasingly significant recent years, possibly due to the emergence of application-to-person (A2P) bulk messaging that allows a large number of SMS messages to be delivered to mobile terminals. It is reported that Americans received 7.4 billion spam text messages during March 2021, a 37% raise compared with February 2020 [4]; also the Federal Trade Commission reveals that in 2020 alone, 2.2 million Americans suffered losses of $3.3 billion to digital fraud and 27% of them are related to SMS spam [16]. The bloom of SMS spam has received attention of Federal Communications Commission and new rules against SMS spam are being considered [6].

**Challenges in fighting against SMS spam**. Fighting against SMS spam is extremely challenging, due to lack of information that enables good understanding of and timely response to emerging spam activities. Although previous research [37, 56] provides insights into spammers' campaigns, spammer's SMS activities and their evading strategies, SMS spam keeps evolving, rendering old knowledge obsolete and protection less effective. What makes things even worse is lack of public, continuously updating, informative SMS spam data collections, which are important for monitoring evolution of spam operations and developing timely and effective countermeasures. To our best knowledge, the two most up-to-date SMS spam datasets are *SMS Spam Collection* [23], and *FBS SMS Spam Dataset* [56]. *SMS Spam Collection* is vastly outdated (donated in 2012), only containing 747 spam messages mostly collected from a UK forum (www.grumbletext.co.uk) which is no longer alive since Oct 2012. *FBS SMS Spam Dataset* was collected from fake-base-station (FBS) messages in China, using a proprietary security app, which makes this dataset not extensible. It is also known that collection of a public, large-scale SMS spam dataset is challenging [56].

In our research, we discovered a new channel that makes it possible to continuously collect high-quality, up-to-date, and also public SMS spam messages, which will greatly facilitate the research and technique development in understanding and mitigating this long-standing and ever-growing security risk. This channel is Twitter on which spam messages are found to be posted by their recipients, for the purpose of seeking advice, warning the public, and notifying the parties impersonated by spammers through smishing (e.g., Amazon), as illustrated in Figure 1. These reported messages are mostly accurate and up-to-date, and come with an increasingly large volume: our research shows that in the past 4 years, the number of posted spam messages grows from around 500 in Q1 2018 to more than 1,700 in Q4 2021 (see Figure 3). So they have become valuable assets for analyzing and detecting SMS spam.

---
*Corresponding author
[†]Also with  School of Cyber Security, University of Chinese Academy of Sciences.
[‡]Also with  Beijing Academy of Artificial Intelligence.

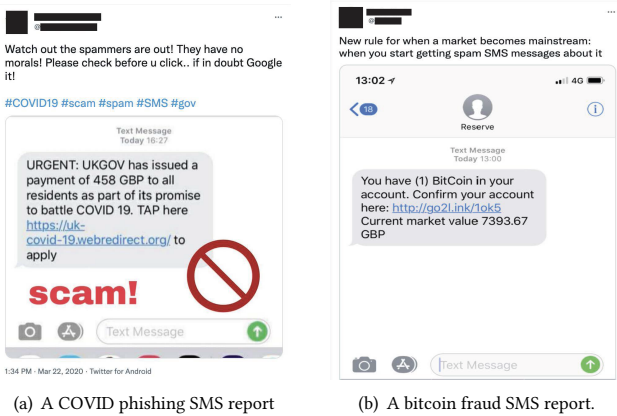

(a) A COVID phishing SMS report

(b) A bitcoin fraud SMS report.

**Figure 1: Typical spam-reporting tweets.**

**Hunting SMS spam from tweets**. To identify the reported SMS spam and extract their content, we developed a new technique to automatically identify the tweets reporting SMS spam and accurately recover from its image attachments spam messages. Our approach, called *SpamHunter*, runs a pipeline that first uses a set of keywords to collect tweets through Tweet APIs, then filters these tweets with image object detection to identify those including SMS screenshots (particularly, an SMS dialog box or text cell), and finally classifies the tweets with SMS screenshots as spam-reporting or not, using a natural language processing (NLP) and machine learning (ML) model. These confirmed spam-reporting tweets are further inspected to extract message content from the attached SMS screenshots, by intersecting the SMS text cell with the text paragraphs captured using a Google Vision API. Our research shows that this pipeline achieves a precision of 95% and a recall of 87%. Applying *SpamHunter* on tweets posted between Jan 2018 and Dec 2021, we discovered 21,918 SMS spam messages in 75 languages. These messages constitute the largest public SMS-spam dataset, which has no overlap with the SMS datasets released before and has never been analyzed. For example, it contains 9,149 spam SMS messages in English so far, much more than *SMS Spam Collection* (with only 747 spam messages) [23]. Meanwhile, the dataset is also diverse (with most messages posted by different Twitter users, see Section 4.1), of high-quality (missed by state-of-the-art SMS spam detectors, see Section 5), and most importantly ever expanding: we built a website [18] to publish and continuously update the dataset.

**Measurement and findings**. On the spam text messages discovered, we performed a measurement study that sheds new light on strategies, targets and infrastructures of today's SMS spam. We found that unlike the SMS-spam data released before [43, 56], which contains mostly less harmful advertisements, most SMS messages reported on Twitter are fraud-related, using fake account alerts, fabricated delivery information, and other tricks like leveraging COVID-19 pandemic (contact tracking, vaccine appointment, etc.) to defraud the message recipients. Also interestingly, reported SMS spam exhibits geographic features: for example, loan and gamble advertisements are pervasive in Indonesia, while credit/debit card scams are reported more frequently in Dutch. Further, our analysis on the URLs carried by spam messages shows that related phishing or malicious websites adopt multiple hosting options including bullet-proof hosting services (e.g., shinjiru), port forwarding services (e.g., ngrok.io), dynamic DNS services (e.g., duckdns.org) and anycast IPs, indicating well-thought-out and well-funded organization of the cybercrimes. Another observation is that 15.4% spam URLs were reported by tweets at least 1 week earlier than they were submitted to VirusTotal (VT), implying that Twitter reports can help block SMS spam more timely than the open threat exchange (OTX) platform. Also, the wide coverage of our dataset makes it possible to find spam campaigns across languages. We detected 53 active multi-lingual spam campaigns, 19% of the 280 SMS campaigns we discovered. These campaigns disseminated similar messages using several languages, e.g., Indonesian and Malay, English and French. The existence of such cross-language campaigns indicates that an organized spam operation might take place in different regions so more effort needs to be made to understand how it works and how to use the information across region to control the risk.

Our measurement on Twitter as a spam reporting system reveals the spam-reporting users' behaviors and the response to their actions, which have never been investigated before. Our study shows that 90% of those users report only once and most of them are rather active on Twitter. Also they tend to tag the organization being impersonated by the spammer (which we call *victim service*) or law enforcement, but often (71% of the reports) fail to get any response from them. Among the popular victim services are banks (e.g., Rabobank), followed by e-commerce companies (e.g., Amazon), postal services (e.g., PostNL), tax authorities (e.g., India Income Tax Department), crypto wallet companies (e.g., Ledger), etc. Surprisingly, most of the targets are local or national, instead of international organizations, which could be due to the spammer's targeting of local businesses or inadequate spam countermeasures from local telecommunication providers. Our further analysis shows that spam activities targeting at the popular victim services usually last for a long time and exhibit a periodic evolution pattern.

Further, we studied whether today's SMS services and text messaging apps offer adequate protection. For this purpose, we sent spam messages through the APIs provided by bulk SMS services (with their consent) and text messaging apps to smartphones under our control. Our automated analysis (on bulk SMS services) and manual inspection (on text messaging apps) show that neither the bulk SMS services nor popular text messaging apps can effectively detect these reported spam text messages, indicating that today's SMS spam detection is still inadequate. We discussed the potential to leverage our new spam dataset to better mitigate the spam risk.

**Contributions**. Compared with prior studies [23, 43, 56], our work provides broad, high-quality and continuously updating information about SMS spam. By the end of 2021, our dataset already includes 21,918 SMS spam messages reported by 14,785 unique Twitter users in 75 languages over the past 4 years.  Following we summarize our contributions:

• *New techniques*. We developed a new framework, *SpamHunter*, that for the first time utilizes social network reports for continuously collecting high-quality and up-to-date SMS spam messages.

• *New dataset*. *SpamHunter* has led to the discovery of tens of thousands of spam text messages, which is made publicly available on

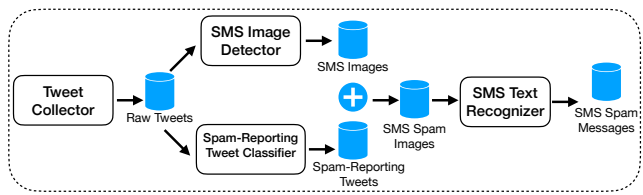

Figure 2: The pipeline of *SpamHunter*.

sites.google.com/view/twitterspamsms. We will periodically update the dataset with newly identified SMS spam cases.

• *New findings.* Upon the SMS spam dataset, we gained new understandings of the criminal activities, including new categories (e.g., COVID-19 related phishing SMS), infrastructures (bullet-proof hosting), and campaigns (cross-language spam campaigns). We also studied the distribution of spam-reporting users on Twitter, and their tagging behavior to understand victim services, which can help better mitigate the spam threat.

## 2 SMS SPAM HUNTER

As mentioned earlier, up-to-date SMS spam data is critical for detecting the ever-evolving malicious activity [46, 52, 54] and for analyzing its operations. However, the existing public spam message datasets [23, 56] are either out-of-date or not extendable, along with other limitations as discussed in §6. In our research, for the first time, we explored a new methodology to continuously scrape reported spam text messages from Twitter, based upon our observation that SMS spam recipients tend to post the text messages on social networks to alert their followers or the organizations impersonated by the spammers, or seek help from relevant parties such as law enforcement agents, as the examples in Figure 1. These spam-reporting tweets often include the screenshots of the spam messages, which we collected using a pipeline called *SpamHunter*.

Although spam-reporting tweets are often characterized by hashtags like #SMS or #spam, direct using them to find relevant tweets does not work well, with a large false positive (unrelated posts or those without screenshots). Further recovering a message from a screenshot is impeded by the presence of noise (e.g., Figure 1(a)). Following we elaborate the design of *SpamHunter*, as shown in Figure 2: our approach first runs a tweet collector to find likely spam-reporting tweets through keywords; followed by an SMS detector to detect those tweets with SMS screenshots attached, and by a tweet classifier to distinguish tweets that complains about SMS spam; in the end, the tweets both complaining spam (determined by the tweet classifier) and attaching messages (by the SMS detector) are further processed through a text recognizer to extract the message content.

**Tweet collector**. Our pipeline starts from collecting users' tweets. For this purpose, we utilized the Twitter Academic API [1], which allows us to search for tweets using complicated query terms. We then composed a query as below to find spam-related tweets, containing a set of keywords such as *spam SMS*, *phishing SMS*, and *scam SMS*. Also, we constrained our search to the tweets carrying at least one image since users tend to attach screenshots instead of pasting the full content when reporting a spam message. In total, the collector found 40,998 tweets with 50,545 image attachments between Jan 1, 2018 and Dec 31, 2021.

(**malicious** OR **spam** OR **phish** OR **phishing**
OR **smish** OR **scam** OR **fraud**) **sms**
has: images

**SMS image detector (SID)**. From the tweets collected, our approach further identifies those including the screenshots of SMS messages. As shown in Figure 1, an SMS screenshot is characterized by a set of unique features, including the sender icon, the date string, the replying box, and most importantly, a dialog box containing message content (SMS text cell). Also, the text cell is usually of a regular shape (e.g., a rounded rectangle), filled with the background color distinct from that of its surroundings. This allows us to utilize Yolov3 [44], an efficient and accurate object detection algorithm, to build the SMS image detector. Our detector recognizes SMS text cells (objects) inside the screenshots we collected and outputs their bounding boxes on the images. To train and evaluate such a detector, we manually labeled images using a YOLO labelling tool, YOLO_Label [2]. Our ground truth set consists of 1000 randomly sampled images, including 687 SMS images with 869 SMS text objects and 313 non-SMS images.

We evaluated the effectiveness of SID, in terms of object classification and localization accuracy. For this purpose, we performed a 5-fold cross validation and the detector found SMS boxes with a precision of 98% and a recall of 97%. To measure the localization accuracy of bounding boxes, we used Interaction over Union (IoU) as the metric: $IoU = \frac{Area_{detect} \cap Area_{object}}{Area_{detect} \cup Area_{object}}$. And SID achieves an IoU of 81%. We then applied it to all the collected 50,545 images, which removed 25,714 non-SMS images and identified 24,831 SMS images with each containing at least one SMS text cell. Our manual inspection on 1000 randomly sampled instances (both positives and negatives) shows a precision of 98% and a recall of 93%. Our detector also outputs the coordinates of each text cell inside its image, which helps recognize SMS text from images later.

**Spam-reporting tweet classifier (SRTC)**. As mentioned earlier, the tweets gathered by the collector may not be related to spam-reporting. For example, some tweets are promotions from anti-spam services. We randomly sampled 500 tweets and found the non-spam-reporting tweets take up to 18%. So *SpamHunter* runs SRTC to remove such noise from our dataset. Spam reporting tweets tend to have negative sentiments, such as unhappiness from the person posting such a tweet. Also, we focus on the tweets with the messages confirmed to be spam by their recipients, instead of those the recipients are not sure about. So in our research, we trained a simple sentiment model (3-layer neural network) [32] to determine whether a tweet indeed reports spam text messages and further ran the classifier to filter out non-spam-reporting tweets.

To train SRTC, we randomly sampled 500 spam-reporting tweets and 250 non-spam-reporting tweets and utilized oversampling [13] to balance the dataset by counting the non-spam-reporting cases twice. Then we translated non-English tweets into English through Google Translation [8]. In a 5-fold cross validation, SRTC achieves a precision of 89% and a recall of 93%.

---
[1]https://developer.twitter.com/en/products/twitter-api/academic-research

[2]https://github.com/developer0hye/Yolo_Label

Table 1: Performance evaluation of *SpamHunter* pipeline.

| Module | Cross-validate (balanced) | | | Test (Inbalanced) | | |
|---|---|---|---|---|---|---|
| | Acc. | Prec. | Recall | Acc. | Prec. | Recall |
| SID | 0.97 | 0.98 | 0.97 | 0.95 | 0.98 | 0.93 |
| SRTC | 0.90 | 0.89 | 0.93 | 0.88 | 0.95 | 0.90 |
| only SID | N/A | N/A | N/A | 0.91 | 0.90 | 0.94 |
| only SRTC | N/A | N/A | N/A | 0.58 | 0.56 | 0.93 |
| SID + SRTC | N/A | N/A | N/A | 0.91 | 0.95 | 0.87 |

**SMS text recognizer (STR)**. The outputs of SRTC and SID are then compared to find those considered to be spam-reporting and also including SMS screenshots. From such selected tweets, STR extracts the content of their messages. This step cannot be done directly using optical character recognition (OCR), as discovered in our research, due to the presence of noise, such as metadata like the timestamp, the sender ID, or phone numbers, replies and user comments, as the examples in Figure 1. Also, many tweets pack multiple messages in a single screenshot. In some cases, both the screenshot of a message and that of the website the SMS points to are displayed together on an image. So our approach utilized the coordinates of each SMS text cell produced by the SID to find the text the cell covers, which is most likely to be the content of a message. For this purpose, we applied the Text Documentation Detection API [7] provided by Google Vision to each SMS image, which recovers text paragraphs along with their coordinates on the image. Then, the coordinates of each paragraph, as bounded by a "paragraph box" on the image, are compared with those of an identified text cell; it is considered to be part of an SMS message if most of the paragraph is within the text cell. Specifically, in our research, we defined $OverlapRatio = \frac{Area_{\text{text paragraph}} \cap Area_{\text{SMS object}}}{Area_{\text{text paragraph}}}$ and set $OverlapRatio = 0.75$ as the threshold to decide whether a text paragraph is part of the given SMS object.

To verify the performance of STR, we randomly sampled 1,000 SMS images analyzed by STR and manually identified the complete content of SMS messages from their screenshots. We then evaluated the text recovered in terms of *word accuracy* and *character accuracy*, where the former is the ratio of correctly recognized SMS words to all identified words, and the latter is the ratio of correctly identified characters to all detected characters. STR achieves a *word accuracy* of 99% and a *character accuracy* of 98%. And it turns out be be much better than running OCR directly on the whole screenshot, which achieves a word accuracy of only 42%. We also measured the accuracy of STR in terms of the whole SMS text recovery, i.e., whether the recovered content of an SMS message is identical to its original text except for special characters. On randomly sampled 1000 SMS images, STR achieves an accuracy of 90%, i.e. accurately recognizing 90% of messages within these images. Among the 105 messages that STR failed to recover their exact message content, we found that SRT missed or added, for 52 messages, no more than 3 words (out of typically more than 20 words in a message), and only deviated from the original text by 6 words or more for 15 messages.

**End-to-end performance evaluation**. Table 1 summarizes the performance of our *SpamHunter* pipeline. We evaluated the performance of each module and the whole pipeline on two datasets: a manually crafted, balanced dataset for training and cross-validation, and a randomly sampled, unbalanced dataset for testing. Each dataset contains 1000 cases and we performed 5-fold cross-validation by default. During the experiment, we noticed that the accuracy of SID and SRTC on the test set are just slightly below those on the cross-validation dataset, which indicates the generality of our model on the whole dataset. Overall, the pipeline achieves a good performance with a recall of 87% and a high precision of 95%. We also compared the performance of the whole pipeline with that of running SID or SRTC only. SRTC alone achieves a poor performance since it failed to distinguish the tweets without SMS images from those with the images. SID alone achieves a precision of 90%, because our keyword-based collector filtered out most irrelevant tweets, while SRTC further boosts the precision to 95%. More importantly, SRTC makes the whole pipeline more reliable, even in the presence of tweets with attached SMS images and spam-related keywords but not spam-reporting, which makes *SpamHunter* more robust against noise.

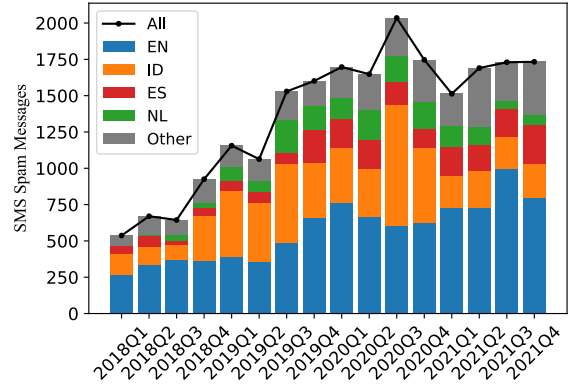

Figure 3: Time trends of SMS spam messages across top languages.

## 3 UNDERSTANDING SMS SPAM

In this section, we report our measurement study on the SMS spam captured by *SpamHunter* (§2). Our research sheds light on the spammer's strategies, based upon the content of spam messages (§3.1), their infrastructures (phone numbers, IP, domain names, etc., see §3.2), as well as the identified spam campaigns (§3.3).

### 3.1 SMS Spam Content

**Landscape**. Through running *SpamHunter* (§2) to scan tweets posted between 2018-01-01 and 2021-12-31, we identified 21,918 unique spam messages from 19,214 spam-reporting tweets. For SMS spam messages reported multiple times, we only count them once and take the date of the first reporting tweet as the reporting time. These messages are in 75 languages, with the top 4 being English (EN, 42%), Indonesian (ID, 25%), Spanish (ES, 10%), and Dutch (NL, 8%), which indicates the pervasiveness of SMS spam activities in these countries and regions. Figure 3 shows the distribution of SMS spam messages over time and languages. As we can see, spam messages reported on Twitter have increased significantly in the recent 4 years, starting from 537 in Q1 2018 to 1,733 in Q4 2021.

The quarter with the most SMS spam messages reported is Q3 2020, mostly due to a huge increase of SMS spam in Indonesian, yet the reason is not clear.

**Categories.** To profile spam categories, we randomly sampled 1,000 out of all the discovered SMS spam messages and manually excluded 53 false-positive cases. We then labelled the remaining 947 spam messages based upon their content. As shown in Table 2, we first defined two main SMS spam categories, *Fraud* and *Ads*, and further refined the labels into 12 subcategories (e.g., *Account alert (Fraud)*, *Promotion (Ads)*). The detailed category definitions and example messages can be found in Appendix A. The labelling process was performed by three researchers independently. When a conflict happened, a further discussion was conducted to reach a consensus.

Among the sampled SMS spam messages, *fraudulent* messages account for 62%. This is because users tend to report the SMS spam considered to be more damaging. The most popular subcategory is *Account alerts (Fraud)* (24.60%), e.g., "*CaixaBank: We regret to inform you that your account has been deactivated, for your security, we ask you to complete the following verification: https://bit.ly/3n7udom*", which is followed by *Promotion (Ads)* (19.75%). Also, we identified emerging and previously unknown spam subcategories. Particularly, 1.69% spam messages are classified to *COVID-19 (Fraud)*, e.g. "*REGISTER FOR COVID-VACCINE from age 18+ Register for vaccine using COVID-19 app. Download from below. Link:http://tiny.cc/COVID-VACCINE*". Also, we observed that 2.39% of the SMS spam messages serve political purposes (in the Politics subcategory). For example, during the 2020 United States presidential election, SMS spam messages were found to either promote or demote some candidates: "*They'll attack your homes if Joe's elected. Pres Trump needs you to become a Diamond Club Member. Your name is MISSING. Donate: bit.ly/3ipuQPr*". Also, *Tax refund (Fraud)* messages never miss any tax return seasons regardless of countries. Here is an example for Australia: "*Due to natural disasters, Australians are entitled to an 8% bonus on their 2020 tax return. Please begin the process by filling out the form below. https://my.gov.verification digital.com*". Another example is for India: "*Attention Taxpayer! Last 11 days left for filing Income Tax Return. File now to Avoid Penalty of Rs 5000 through All India ITR App www.gs.im/VgGFEpr1TLAB*". More examples in the categories can be found in Appendix 12.

**Distribution of SMS spam over languages.** Also, we observed that the distribution of SMS messages over those subcategories varies in different languages. *Account alert (Fraud)* turns out to be the most common spam subcategories for most languages except Indonesian, in which *Promotion (Ads)* messages are more popular. Specific spam subcategory can receive much more attention (and complaints) in one language than another languages. For example, *Loan/Gamble (Ads)* messages account for 36% spam messages in Indonesian, and are few reported in other languages; 29% of SMS messages in Dutch are related to *Credit/Debit card (Fraud)*, far more than the ratio in other languages. This may be because that the loan and gambling industry are popular in Indonesian areas, and banks in Dutch are more likely to be the target of the credit/debit card fraud activities. Such findings can help understand the SMS spam ecosystem in specific language, which could help take more targeted mitigation countermeasures.

Table 2: Reported spam message categories.

| Spam Category | Subcategory | Spam ratio | # Labelled messages |
|---|---|---|---|
| Fraud | Account alert | 24.60% | 233 |
| | Finance | 10.14% | 96 |
| | Prize | 8.24% | 78 |
| | Delivery | 6.12% | 58 |
| | Credit/Debit card | 5.49% | 52 |
| | Tax refund | 2.53% | 24 |
| | COVID-19 | 1.69% | 16 |
| | Other | 3.48% | 33 |
| Ads | Promotion | 19.75% | 187 |
| | Loan/Gamble | 9.40% | 89 |
| | Politics | 2.53% | 24 |
| | Other | 6.02% | 57 |

Table 3: VirusTotal reports for spam URLs and FQDNs: Here VT-M is the VT category of malicious, VT-MW is the VT category of malware and VT-P is the VT category of phishing.

| Category | Num | VT ≥ 1 | VT ≥ 5 | VT-M | VT-MW | VT-P |
|---|---|---|---|---|---|---|
| Spam URLs | 12,455 | 16.53% | 6.50% | 11.77% | 5.64% | 12.42% |
| Spam FQDNs | 8,313 | 20.75% | 8.23% | 16.17% | 7.99% | 15.59% |

**Spam URLs.** Among the 21,918 SMS spam messages identified, 14,185 (64.72%) contain at least one URL, with 12,855 unique URLs and 7,093 fully qualified domain names (FQDNs) in total. We observed that URL shortening services, such as *Bitly* and *s.id*, are widely used to generate spam URLs. Particularly, the top 20 URL shorteners (see Appendix 14) account for 27.82% of all spam URLs. For each shortened URL, we tried to recover its final landing URL through visiting each shortened URL and recording its redirection chain. Altogether, we obtained a collection of 12,455 unique recovered spam URLs and 8,313 FQDNs.

We further analyzed the spam URLs and FQDNs discovered by *SpamHunter*, leveraging the most prominent open threat exchange (OTX) platform – VirusTotal (VT). We found VT reports for 39.54% of the 12,455 spam URLs, which include the detection results generated by 83 AV engines (e.g., blacklists) [21]. The more AV engines that flag a URL/FQDN, the more likely the URL/FQDN is malicious [21]. Here we denote the set of URLs/FQDNs detected by at least 1 AV engine as $VT \geq 1$ and the set of those flagged by at least 5 engines as $VT \geq 5$. We consider $VT \geq 5$ since it is a common threshold for determining whether a spam URL/FQDN is malicious [40, 41]. Also on a report is coarse-grained categories which VT assigns to each flagged URL/FQDN, i.e., *malicious*, *malware*, or *phishing*. Note that a URL/FQDN can be classified into multiple categories. Table 3 illustrates the distribution of spam URLs and FQDNs over $VT \geq 1$, $VT \geq 5$ and those VT-specified categories. Among all the spam URLs we discovered, 16.53% were flagged by at least one AV engine ($VT \geq 1$) while 6.50% were detected by at least 5 AV engines ($VT \geq 5$). Compared with spam URLs, spam FQDNs are a bit more likely to be flagged by AV engines, e.g., 8.23% spam FQDNs are detected by at least 5 AV engines ($VT \geq 5$) compared with 6.50% spam URLs in $VT \geq 5$.

**Timeliness of open threat intelligence platforms.** We also profiled how timely OTXs can raise an alarm on a spam URL: that is, how quickly OTXs can flag a spam URL after a spam campaign starts.

We answered this question by profiling for each spam URL the time gap, denoted by $Date_{VT} - Date_{Twitter}$, where $Date_{VT}$ means the date when VT scanned the URL and flagged it as harmful for the first time, and $Date_{Twitter}$ represents the date when it was first reported on Twitter. The larger the time gap, the further the VT report is lagged behind the spam reporting on Twitter. For the spam URLs in $VT \geq 1$, 30.1% of them were detected later by 1 day on VT compared with their first Twitter reporting dates. And 15.4% of the spam URLs were lagged longer than 1 week. This indicates that our *SpamHunter* can help OTXs identify spam URLs more timely, possibly because the reported URLs appear in screenshots and therefore may not be immediately captured by search engines. For example, an SMS phishing URL targeting customers of the HSBC bank was complained as early as March 2018, but first scanned and detected by VT on Sep 17, 2019 (564 days later). In another case, a Twitter user complained about a phishing attack targeting Apple ID theft on Oct 27, 2018, but VT did not flag the reported phishing URL (http://apple-expiry.com) until 145 days later.

### 3.2 SMS Spam Infrastructure

**Sender Phone numbers**. We noticed that many spam-report tweets expose the sender's phone number or ID (e.g., brand name) in their attached SMS screenshots. For example, the SMS screenshot in Figure 1(b) shows that the reported SMS comes from a sender ID *Reserve*. Based on the observation that the sender's phone number or ID tends to be well-formatted and appears on top of SMS screenshots, we extracted such information by converting each SMS screenshot into texts and performing a pattern matching on the text above SMS messages. Manual verification reveals that the heuristic extraction method can correctly recognize the phone number or ID for 99 out of the 100 randomly sampled screenshots.

Using this approach, we have extracted 9,092 sender phone numbers or IDs from the 21,563 SMS images. Among them, 6,895 are regular phone numbers, 843 are short codes, and 1,354 are sender IDs. We further queried the Twilio Lookup API [20] for WHOIS information of these numbers, among which 6,878 (6035 regular phone numbers and 843 short codes) contain a valid WHOIS information identified and returned. Figure 4 illustrates the countries and telecommunication providers associated with most phone numbers. As we can see, Indonesia, Netherlands, and India are among the top 3 countries where these sender phone numbers belong to, and the top 3 telecommunication providers are *Telkomsel*, an Indonesian carrier, *Vodafone*, a UK carrier that also has a large number of customers in India, and *KPN* from Netherlands. The results are consistent with our analysis on the language distribution in S3.1, which also implies the popularity of SMS spam in Indonesia, Netherlands, and India. Results from the Twilio API also show that 97% of the sender numbers are mobile numbers, 2% are VoIP numbers, and the remaining 1% are landline numbers that we believe are misclassified by Twilio, as pointed out by the prior research [43].

**Blacklisted Phone numbers**. To understand whether these identified phone numbers have been blocked or blacklisted, we queried two popular and well-known spam call block services, i.e. *Showcaller* [9] and *Robokiller* [19]. Both services provide APIs that take a phone number as input and returns whether it is spam (suspicious) or neural. According to their documentation, *Showcaller* maintains phone number blacklists for 6 countries, including United States, Canada, United Kingdom, India, Australia and Singapore, and *Robokiller* only works on United States/Canada phone numbers. We then utilized the carrier information to filter phone numbers in the supported countries. Among 4,688 spam phone numbers identified in the 6 qualified countries, *Showcaller* detected only 36 of them as spam or suspicious, and *Robokiller* only reported 8 as spam among the 255 United States/Canada spam phone numbers. To conclude, both spam call blockers have a very low detection rate (less than 5%).

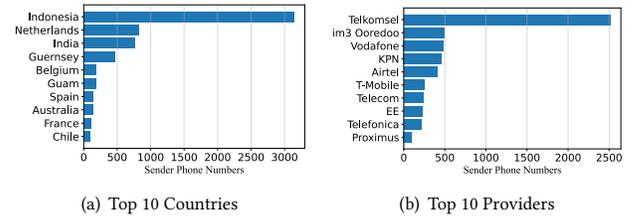

(a) Top 10 Countries    (b) Top 10 Providers

**Figure 4: Top 10 countries and carriers of spam sender phone numbers.**

A problem here is the difficulty in excluding spoofed phone numbers: the spammer can use fake caller IDs (e.g., phone number, and sender ID) to conceal the real source of a SMS spam message. The presence of such numbers may add noise to those extracted phone numbers and affect our measurement results. Phone number spoofing can be carried out through many channels such as fake base stations [33, 56], and VOIP [26, 53]. However, existing spoofing techniques do not work well on mobile numbers, which constitute the majority of our dataset (97%): spoofing through fake base stations usually only works in the GSM (2G) cellular network due to its vulnerable authentication mechanisms [33, 48]. Therefore, we believe that the identified spam phone numbers are mostly valid. On the other hand, spammers may purchase temporary mobile phone numbers from resellers like Twilio. Thus the distribution of spam phone numbers (e.g., countries, carriers) may not reflect the real distribution of spammers.

**Network infrastructures**. We profiled the network infrastructure of SMS spam activities. From passive DNS datasets [2], we identified 33,495 IPs that the 8,313 spam FQDNs were resolved to during their lifetimes. Among these IPs, 5,751 have hosted the spam FQDNs flagged by VirusTotal ($VT \geq 1$), while 1,142 are associated with the spam FQDNs reported by at least 5 detection engines ($VT \geq 5$). Appendix 15 shows their distribution in terms of countries, autonomous systems (ASes), and network blocks, as extracted from IPinfo [10]. We can see that the network infrastructure of spam activities is widely distributed across countries and ASes.

We further looked into the WHOIS information of these spam IPs, aiming to uncover the underlying hosting providers potentially abused by these spammers. We consider spam IPs associated with spam FQDNs in the set of $VT \geq 5$ as *malicious-related* IPs, which is a commonly used threshold [40, 41]. The detailed information is listed in Appendix 16. Although many of these spam IPs are registered under popular cloud providers especially Amazon (39%) and Google (6%), these providers' IPs are less likely to be flagged as malicious-related, e.g., Amazon only accounts for 20% spam IPs in the set of $VT \geq 5$. Meanwhile, smaller hosting providers' IPs are more likely to be flagged as malicious-related, e.g., 20% of the spam IPs in $VT \geq 5$ belong to Namecheap while it only contains 2.0% IPs among all the IPs we identified. Other examples include *Shinjiru* (a small company dedicated to offshore web hosting), and *DigitalOcean*. One interesting observation is that Namecheap has been tagged many times by Twitter users because of the spam websites hosted on it.

**New trends**. An trend we observed is the *increasing* adoption of port forwarding services and dynamic DNS services by spammers to hide their network infrastructure. Traditionally, the operators of spam activities involving websites need to register domain names and purchase hosting IP addresses and servers. This renders spammers hard to migrate to new infrastructure. Even after they mitigated, the spam domains and IPs may still contain the evidence of illicit activities, which can help trace and block those spam activities. However, among the spam URLs observed in our dataset, 176 (from 185 spam URLs) are found to belong to Ngrok [11], in the form of XXX.ngrok.io/XX, e.g., https://5d6986714c90.ngrok.io/sbibank/, and https://09a62a22.ngrok.io/. Ngrok is a popular Internet-wide port forwarding service (PFS) and can transparently forward the network traffic toward a specific Ngrok subdomain (XXX.ngrok.io) to the IP and port designated by its customers. By abusing PFSes, spam operators no longer need to register domain names or possess any pubic IP address, and thus can hide the spam infrastructures and evade blocking. Our research indicates that such PFS-facilitated spam activities are on the rise: from 2019 to 2021, the number of related spam URLs grows from 14 to 139 in our dataset.

Also, 52 subdomains (e.g., hgjjcchslo.duckdns.org) of a dynamic DNS service, duckdns.org, are found from 78 spam messages for automatic DNS updating, which not only allows spammers to operate spam websites without registering domains, but also facilitates them to quickly migrate to new network infrastructure. Similar to PFS, such malicious activities grow quickly these years, evidenced by doubling of spam subdomains of duckdns.org from 2020 to 2021. To our best knowledge, we are the first to report the abuse of these services in SMS spam activities. We also identified the use of anycast IPs in spam web hosting. Traditionally, anycast IPs serve the purpose of addressing DNS servers such as 8.8.8.8 operated by Google DNS. However, we observed 2,564 anycast IPs for addressing spam URLs and domains, which accounts for 7.65% of all spam IPs. For example, 67.199.248.11 served 1,551 spam messages and 1,168 spam URLs while 99.83.154.118 was associated with 191 spam messages and 185 spam URLs (e.g., http://scotia-bank-support.com and http://security-payee-check.com).

### 3.3 SMS Spam Campaigns

On the SMS spam messages identified, we studied the underlying correlations. We first clustered SMS spam messages into groups based on their URLs and then profiled the *spam campaigns*.

**Spam clustering**. Previous research [31, 35, 56] on spam campaigns clusters spam messages based on URL, text similarity, or contact information. On our dataset, we utilized spam URLs to cluster SMS spam messages. Specifically, we put each message in a separate cluster and then continued to merge different clusters whenever they share a spam URL among at least one message in each cluster, until the merge could not proceed. Although this clustering strategy is conservative and likely to over-estimate the number of campaigns, i.e., splitting spam messages of the same campaigns into multiple clusters, it still sheds light on the real-word SMS spam campaigns. In total, we have identified 17,670 distinct clusters, with 280 clusters including more than one messages, which are considered as *spam campaigns*.

**Cross-language spam campaigns**. Among these 280 spam campaigns, 53 of them contain SMS spam messages in different languages. For example, a malicious spam URL https://n26-app.com is associated with 6 unique spam messages in 4 different languages, including English, French, German, and Spanish. Figure 6 (see Appendix C) presents the screenshots of four spam messages belonging to the same campaign but in different languages. The presence of these cross-language spam activities has two important implications. First our finding indicates that the spam operator may organize a spam operation across countries, so effective response to the operation may need international cooperation. Second, the artifacts of the same campaign in different languages can potentially help the spam mitigation effort: e.g., a spam message reported by tweets can be translated into other languages to support spam filtering in these languages, even before the spam messages in the languages show up. This purpose could be served by the observation that such related spam messages could be reported at different times. For example, in Figure 6, the tweet reporting the German version of the spam message appeared on Apr 14, 2021, one week earlier than its Spanish version.

## 4 UNDERSTANDING SPAM REPORTING

Following the measurement on SMS spam content ( §3), we move onto spam reporting activities with a focus on spam reporters, their distributions, recipients of the reports and the responses.

### 4.1 Spam Reporters

As mentioned earlier, 21,918 SMS spam messages we discovered are extracted from 19,214 *spam-reporting tweets* (*SRTs*) published between Jan 2018 and Dec 2021. These tweets come from 14,785 different *spam-reporting Twitter users* (*SRTUs*). In our research, we profiled these SRTUs to analyze their behaviors. We found that 90% SRTUs contributed only one SRT while 99% of them reported less than five SRTs. Only five SRTUs posted more than 50 SRTs, and all of them are identified as individual accounts that tend to report their daily spam SMS messages. These 5 SRTUs contributed 1,264 SMS spam messages, accounting for nearly 7% of the SRTs we collected. So we conclude that *spam reporting activities are widely spread across a large number of Twitter users, with most contributing equally, despite the presence of a very few SRTUs with many SRTs*. Through Twitter APIs, we are able to access public metrics of Twitter users, including account lifetime, number of followers, and number of tweets posted. These SRTUs are found to have an average lifetime of 9 years with 93% of them having posted more than 100 tweets, and 83% with followers ranging from 20 to 10,000. These statistics indicate that most SRTUs are fairly active on Twitter.

**Sources & locations**. Twitter APIs also allow us to query the user agent (e.g., Twitter on Android and Twitter Web) for each tweet.

Table 4: Categories of top 128 tagged Twitter accounts.

| Category | Count | Ratio | Examples |
| --- | --- | --- | --- |
| Victim service | 56 | 43.75% | @Paytm |
| Law-enforcement | 35 | 27.34% | @policia |
| Cellular carrier | 18 | 14.06% | @Telkomsel |
| Individual | 7 | 5.47% | @rsprasad |
| Anti-spam service | 5 | 3.91% | @fraudehelpdesk |
| Other | 7 | 5.47% | @namecheap |

Among all SRTs, 73% were posted from Twitter apps on either Android or iOS, which is consistent with our observation that SRTUs tend to take a screenshot of the spam message they received before posting it on Twitter. Another 19% reports were published through the Twitter Web platform while the remaining 8% came from other platforms such as TweetDeck or Tweetbot. We also noticed that some SRTUs enabled coarse-grained location sharing when publishing SRTs, which makes 6% SRTs location-aware. Since these location-aware SRTs can shed light on where the SMS spams were reported, we thus analyzed the distribution of these SRTs across different countries. Specifically, the top 3 countries are India with a share of 24% location-aware SRTs, Indonesian(15%), and Netherlands (11%). Other countries among the top 10 include United Kingdom (8%), Spain (6%), Australia (5%) and United States (3%). The results are consistent with the language distribution of SMS spam messages discussed in §3.1.

### 4.2 Recipients of Spam Reports

**Tagged Twitter accounts**. In our study, we noticed an interesting spam-reporting behavior pattern: Twitter users tend to tag accounts of relevant parties when reporting SMS spam messages. Our further investigation into such behavior reveals that the interactions between spam victims and the related parties can have significant security implications. Specifically, among the 19,214 SRTs, 12,817 (67%) tagged at least one Twitter account while 5,130 (27%) tagged at least two. In total, we identified 5,495 unique Twitter accounts, which we call *tagged accounts*. We further looked into the distribution of the tagged Twitter accounts over the volume of tags they have received, which reveals that most tag activities are associated with a small set of Twitter accounts. Specifically, the top 128 Twitter accounts are associated with 50% of tag activities: e.g., @Telkosel, the largest cellular carrier in India, received 901 tags during our measurement period, while @rabobank, a Dutch bank company, was tagged 376 times for reporting related spam SMS messages.

We then profiled the types of the aforementioned top 128 tagged Twitter accounts, as presented in Table 4. Here we name the target services or companies the SMS fraudster impersonates as *victim service* (e.g. @Paytm, @Rabobank). The accounts of such victim services are the most common targets of tagging, constituting 56 out of the top 128 tagged accounts. Following these accounts are those belonging to law-enforcement and cellular carriers, to which SRTUs also tend to make complaints. Other top tagged Twitter accounts belong to various parties such as political leaders, third-party anti-spam services, and even a web hosting provider (Namecheap). Among the victim services, *banking* services (31) are the most popular targets, followed by *e-commerce* (9), *digital payment* (6), *postal*

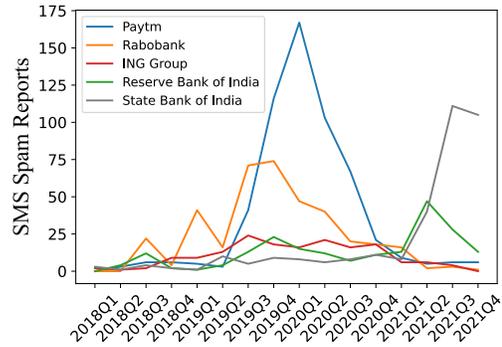

Figure 5: Temporal evolution of SMS spam reports on top 5 victim services.

*service* (4), and *tax authorities* (4). We also noticed two cryptocurrency trade companies, @ledger and @lunomoney, which attackers impersonated in an attempt to steal the credentials of users' cryptocurrency wallet through phishing SMS messages.

**Evolution of SMS spams targeting victim services**. Leveraging SRTUs' tag activities on Twitter, we are able to link spam SMS reports to the victim services spammers impersonate. None of the spam SMS datasets released by prior work provide such information. Given these victim services together with their spam reports, it is interesting to measure the evolution of all the spam campaigns targeting a specific victim service over time. Figure 5 presents the temporal distribution of SRTs tagged the top 5 victim services: Paytm (an online payment service), Rabobank (a banking and financial service), ING Group (a Dutch banking and financial service), Reserve Bank of India (RBI), and State Bank of India (SBI). From the figure, we can see SRTs for each victim service are not distributed evenly across time. Taking Paytm, the victim service associated with most spam reports, as example, spam reports start in Q1 2018 in a low volume, then gradually move upward to the peak during Q1 2020, before moving down the ramp to low intensity again in Q2 2021. Unlike Paytm, spam reports on SBI see a significant sudden increase starting from Q2 2021 after a long time of relatively low intensity. One possible explanation is that spam operators may switch among multiple victim services to maximize their profits and avoid being taken down. We also observed a periodic pattern of crest and trough for the spam reports associated with a specific target. A typical example is Rabobank, whose report number goes up every other quarter between Q1 2018 and Q3 2019 and then slowly goes down since Q1 2020. Regardless of the temporal evolution, spam reports have been received in almost every quarter for all the 5 victim services, which suggests that a long-term effort should be made to fight against SMS spam.

**SMS spam templates**. We also found that SMS spam messages targeting different victim services can share the same message templates, which is quite common in the reported spam messages. From those targeting the top 6 victim services, we identified two common templates. Specifically, an SMS spam message template with an obvious grammatical error, *Dear customer your XXX point[s] worth Rs XXXX expired by XX/XX/XXXX. Kindly convert your points into cash by click here {URL}*, has been reported over 50 times, which targets Kotak Bank Debit Card, SBI Credit Card, and ICICI Saving A/C. Another spam template like *[XXX] Your account has been placed in the quarantine zone due to suspicious login attempts, you can reactivate your account via: https://xxx-quarantinezone.com* has been shared by the messages aiming at both Rabobank and ING Group. The pervasiveness of such template sharing indicates that either the same spam campaign targets multiple victim services or different campaigns share the same text generator.

### 4.3 Response to Spam Reports

We also looked into further interactions between the SRTUs and the tagged Twitter accounts, aiming to profile the effectiveness of spam reporting. Twitter associates each tweet thread with a unique conversation identifier, which records the first tweet id of the whole conversation and thus allows us to extract all replies following a specific tweet. For each spam-reporting tweet, we extracted all the replies within 30 days following the tweet. As aforementioned, 12,817 SRTs have tagged relevant Twitter accounts, among which, 10,134 (53%) received at least one replies. We further matched these replies with the tagged Twitter accounts, and found that only 5,580 (29%) spam-reporting tweets got replies from at least one of the tagged Twitter accounts. Appendix 11 illustrates the reply rate of top tagged Twitter accounts. 51 (40%) tagged accounts never replied to users' SMS reports, including 17 law-enforcement accounts like *Telecom Regulatory Authority of India*, 17 victim services including both cryptocurrency wallet companies *Ledger and Luno*, and 14 other types accounts. Also, 73 (58%) tagged accounts replied less than 30% of users' spam reports. Our measurement results show that most users' SMS spam reports on Twitter do not get timely response from related parties, especially the victim services impersonated by spammers and law-enforcement accounts, which may look into these cases at an early stage and take countermeasure actions such as alerting other users to similar tricks.

## 5 EVALUATING SMS SPAM DETECTION

We evaluated the robustness of real-world anti-spam services and infrastructures, by running our evaluation framework (§5.1). Here we report our findings, focusing on three key players in the SMS ecosystem: third-party anti-spam services such as OOPSpam and Plino, anti-spam protection of bulk SMS providers, and popular text messaging apps.

### 5.1 Evaluation Framework

To understand the effectiveness of the spam messages collected by *SpamHunter*, we investigated whether real-world anti-spam protection could prevent them from reaching targeted recipients. For this purpose, we built an evaluation framework to study how popular anti-spam services, bulk SMS services and text messaging apps react to these messages. In our research, we do not consider anti-spam countermeasures that utilize information such as spam number blacklists and focus on the general content-based SMS spam filtering only.

**Crafting the SMS spam testing set**. To perform the evaluation, we constructed a representative testing set of two groups: (1) the *Twitter-reported* subset including 100 SMS spam messages randomly sampled from the data collected by *SpamHunter*, and (2) the *historical* subset with 50 spam messages randomly sampled from *SMS Spam Collection* [23]. All the messages in the testing set have been manually inspected to remove non-spam messages and to correct OCR errors (including removing extra spaces). As shown in Table 5, the testing set consists of 150 unique SMS spam messages, with exactly 75 *Ads* messages and 75 *Fraud* messages. Since some SMS platforms (e.g., bulk SMS services) may not support non-English messages, we translated all 67 non-English texts in the testing set into English and set the default country to the United States during our evaluation.

Table 5: Categories of the crafted SMS spam testset

| Source | Ads | Fraud | Total |
|---|---|---|---|
| Historical | 34 | 16 | 50 |
| Twitter | 41 | 59 | 100 |
| Total | 75 | 75 | 150 |

**Evaluating anti-spam services**. Third-party anti-spam services typically rely on content, rather than metadata (e.g., the sender's phone number), to detect spam, such as OOPSpam [12] and Plino [15]. Also some text moderation services such as Perspective [14] also support spam detection. These services usually leverage machine learning algorithms to detect spam content but are known to be vulnerable to well-crafted adversarial examples [39]. In our research, we evaluated the performance of popular anti-spam services on our testing set. Specifically, we selected 3 popular anti-spam services, i.e. OOPSpam, Plino, and Perspective. All these services are content-based, for generic spam detection, not just for SMS spam. They provide APIs that accept text as an input and return the detection result: OOPSpam and Plino report a text label (e.g., spam or ham), while Perspective outputs a spam probability score between 0 and 1, for which we set a cutoff of 0.5 to determine whether the input content is spam.

**Evaluating bulk SMS services**. Bulk SMS services (BSSs) offer an convenient, programmable web API to send and manage SMS messages in bulk. Usually, BSSs cooperate closely with multiple cellular carriers to provide service for clients in different regions. To avoid complaints from these cellular carriers, BSSs also equip with a message inspection system to prevent abuse from spammers. However, it is still unknown whether such protection indeed works. Our workflow starts with a generalized SMS message sender that supports interaction with various bulk SMS services through their web APIs. We utilized 4 smart phones with valid SIM cards as the recipient devices. On each phone, we also installed a controller to automatically check the success reception of the hundreds of messages delivered through the bulk services. Since each SMS message can only contain no more than 160 characters and a long message will be broken into multiple messages, we added a unique 6-digit ID to each message to make it easier to be recognized by the controller. The delivery time of a message varies due to the network delay, thus we consider that a bulk SMS service fails to block a given SMS spam message when it has been delivered to any of the 4 recipient devices within 5 minutes. In our research, we explained our testing procedure to top 10 bulk SMS providers [1] and 3 of them (Twilio [20], TextMagic [3], ClickSend [5]) gave us permissions to conduct the experiments. The results are presented in §5.3.

Table 6: Text messaging apps under our evaluation: CSSF denotes content-based SMS spam filtering.

| App | Platform | #Installs | CSSF |
|---|---|---|---|
| Android Message | Android | > 1B | No |
| RoboKiller | Android | > 5M | Yes |
| smsBlocker | Android | > 1M | Yes |
| Antinuisance | Android | > 1M | Yes |
| iOS Message | iOS | - | No |
| SMS Shield | iOS | - | Yes |
| VeroSMS | iOS | - | Yes |

Table 7: Evaluation results for anti-spam services.

| Service | % Detected | Twitter | Historical | Ads | Fraud |
|---|---|---|---|---|---|
| Perspective | 95% | 96/100 | 47/50 | 69/75 | 74/75 |
| OOPSpam | 91% | 91/100 | 45/50 | 67/75 | 69/75 |
| Plino | 79% | 82/100 | 37/50 | 61/75 | 58/75 |

**Evaluating text messaging apps.** Text messaging apps are used to send or receive messages and often provide anti-spam features as well. We selected text messaging apps by considering their popularity, anti-spam features, and whether they were studied by prior research [42]. As listed in Table 6, we selected 7 popular text messaging apps, including the native text messaging apps on Android and iOS, i.e. *Android Message* and *iOS Message*. In our experiments, we used three smartphones (2 Google Pixel 4 and one iPhone 12) for an evaluation test: one of the Google Pixel 4 served as the spam sender while the other two as the receivers. The controller on the sending phone automatically issued SMS-spam messages to the recipient phones. Since text messaging apps have heterogeneous GUIs, we manually checked each text messaging app to decide whether the messaging app had flagged or blocked the received spam messages.

**Ethical considerations**. We took ethical issues seriously and tried our best to minimize the potential side effects in the evaluation. Specifically, we checked our crafted testing set and made sure that all sensitive information, like personal names or bank account numbers, was excluded. We found that most SRTUs had already masked such sensitive information with graffiti or mosaic before posting SMS spam screenshots. As mentioned earlier, we also explained to the bulk SMS service providers in detail about our methodology and goals, and performed the evaluation with their consent.

### 5.2 Anti-Spam Services

We present the findings of our evaluation study on anti-spam services in Table 7. Overall, Perspective and OOPSpam achieved a high detection rate of 95% and 91% respectively. Plino did not perform as well as the other two anti-spam services on both Twitter-reported and historical data. We also compared the results of testing on *Ads* and *Fraud* messages. Except Plino, Perspective and OOPSpam achieved slighter better performance on *Fraud* messages than *Ads* messages. However, they still missed some harmful cases. For example, *"Due to suspicious activity your Apple-id has been LOCKED tap https://apple.id-loginauth.com/ to restore full access to your Apple services"* was mis-classified as benign by Perspective and Plino. Also, OOPSpam missed the case such as *"Sabadell - ES": User disabled for security reasons. www.bit.ly/2BKJ25E activate now*".

Given the high alarm rates from anti-spam services, we further analyzed their mis-alarm rates, in other words, how likely an anti-spam service would falsely flag benign SMS messages. Specifically, we looked into Perspective since it has detected most spam texts in our dataset. We selected 124 benign SMS messages from the ground truth dataset we labeled when evaluating *SpamHunter* (§2), and then fed them to the Perspective API. As a result, it falsely reported 103 out of 124 benign SMS messages. Such falsely detected messages include anti-fraud SMS, e.g. *"Please urgently call Fraud Prevention on 03456031832 from UK or intl +441226260049 24x7 quoting reference cd. This is not a marketing text. Do not reply by SMS"*, OTP (One-time password) SMS, e.g. *"Use 625546 for two-factor authentication on Facebook"*, and normal service SMS, e.g. *"Hi! We are happy to inform you that your IKEA order XXXXXXXXX has been delivered. If you have any further questions or need additional help - please contact customer service at 1-888-888-4532"*. These normal messages are quite common in the real world. Our results indicate that anti-spam services may not be very suitable for real-world SMS spam detection since such a high false positive rate will generate overwhelming false alarms and significantly undermine the business of the parties like bulk SMS services and cellular carriers.

### 5.3 Bulk SMS Service

As described in §5.1, we evaluated the robustness of the anti-spam protection deployed by 3 popular bulk SMS services.

**Challenges in evaluating bulk SMS services.** Evaluating bulk SMS services turns out to be more difficult than expected. Different from anti-spam services, bulk SMS service providers deploy a complex detection system, which is a black box to their customers. This forces us to rely on our recipient devices, based upon whether they receive issued messages, to determine whether the spam texts have been detected. On the other hand, a testing message may fail to be delivered due to other reasons such as network errors or being blocked by the carrier network. Thus the findings made in our evaluation test are just an upper bound for the performance of bulk SMS providers' anti-spam protection. Another problem for the evaluation study comes from the user account, which can be suspended or even terminated by bulk SMS providers for various reasons such as reaching the quota of the messages allowed to be sent within a time window. As a result, the user will be required to provide information for incident investigation, which can take up to several days. These factors have unfortunately made our evaluation process time-consuming and less scalable. In the end, we were able to distribute all messages in our testing set through all the bulk SMS services except for TextMagic, which stopped our testing after sending 136 messages due to a complaint from some cellular carrier.

As shown in Table 8, all three bulk SMS services have a low spam-blocking rate, ranging from 6% to 17%, on both the spam message groups (Twitter-reported and historical, see Section 5.1), indicating that they could be easily circumvented by SMS spam campaigns. We further looked into the categories of the spam messages blocked by these providers. It turns out that *Fraud* messages are more likely to be blocked by all three bulk SMS providers than *Ads* messages. This makes sense since bulk SMS services are commonly used by companies to promote their products and bulk SMS services have strong motivations to be more tolerant of *Ads* messages. However, more than 70% of the *Fraud* messages still passed the detection and reached our recipient devices. Among these messages, 50 were missed by all three bulk SMS providers. Here is an example: "*Your internet banking has been disabled for security reasons, Please visit your local branch or unlock at http://217.138.118.54*".

Table 8: Evaluation results for bulk SMS services.

| Service | % Blocked | Twitter | Historical | Ads | Fraud |
|---|---|---|---|---|---|
| Twilio | 6% | 6/100 | 3/50 | 3/75 | 6/75 |
| ClickSend | 12% | 12/100 | 6/50 | 8/75 | 10/75 |
| TextMagic | 17% | 19/92 | 3/44 | 4/67 | 18/69 |

Table 9: Spam filtering performance of text messaging apps.

| App | % Alarmed | Twitter | Historical | Ads | Fraud |
|---|---|---|---|---|---|
| Android Message | 0% | 0/100 | 0/50 | 0/75 | 0/75 |
| iOS Message | 0% | 0/100 | 0/50 | 0/75 | 0/75 |
| AntiNuisance | 0% | 0/100 | 0/50 | 0/75 | 0/75 |
| RoboKiller | 48% | 49/100 | 23/50 | 39/75 | 43/75 |
| VeroSMS | 69% | 56/100 | 47/50 | 50/75 | 53/75 |
| SMS Shield | 73% | 63/100 | 46/50 | 57/75 | 52/75 |
| smsBlocker | 73% | 73/100 | 36/50 | 49/75 | 60/75 |

**Feasibility of abusing bulk SMS service.** To evaluate the feasibility of abusing bulk SMS services for spamming, we also studied the background check performed by the 3 bulk SMS providers to vet their users. All of them provide a free trial with only an email and/or phone number needed to register an account. During the trial period, ClickSend and TextMagic allow the users to send free SMS messages within a trial budget. To abuse such services, an attacker can register many trial accounts to send out spam messages. On the other side, Twilio has more restrictions on free trial accounts, e.g., SMS messages can only be sent to manually verified phone numbers during trial. And an attacker has to deposit a minimal of $20 before sending SMS, which raises the bar for abuse.

**Responsible disclosure**. As mentioned above, we shared our evaluation results with the bulk SMS service providers at our best efforts. So far, we have not received any response from ClickSend and Textmagic, and Twillio replied us that they will keep an eye on our detection results.

### 5.4 Text Messaging Apps

As described in §5.1, we selected 7 popular text messaging apps with anti-spam protection. We then evaluated them on the crafted testing set. Table 9 presents our evaluation results. Both native SMS apps on Android and iOS, i.e. Android Messaging and iOS Messaging, failed to detect any of the 150 testing messages. This is because they all take a conservative strategy and only block the messages from specific phone numbers. although AntiNuisance is featured with anti-spam capability, it also failed to capture any spam message, probably due to its reliance on keyword-based filtering. An interesting observation is that another two anti-spam texting apps, SMS Shield and VeroSMS both achieved over 90% detection rate on historical spam messages but could only catch less than 70% of the spam messages reported on Twitter. This indicates that their anti-spam models do not adapt well to the evolution of spam messages. On the *Fraud* messages, smsBlocker has the highest detection rate, i.e. 80%, yet still misses 15 malicious messages, e.g., "*KNAB: We have placed your account in the "quarantine zone". You must confirm your device before 04-09 to avoid blockage: sg9.top/NL-KNAB*".

We also measured the mis-alarm rate of the top 3 messaging apps, i.e., VeroSMS, SMS Shield and smsBlocker, on the 124 benign messages aforementioned. The mis-alarm rates of these messaging apps range from 38% to 46%. Compared with anti-spam services, messaging apps achieve a lower mis-alarm rate. However, they still falsely flagged some normal service messages, e.g., "*Out for Delivery (Replacement): ... tracking ID XX from flipkart.com will be delivered on successful pickup verification, today by an EKART Wish Master (call XXXXXXXXXX, PIN XXX). Keep the product ready with all accessories and tags for verification*" was flagged as spam by smsBlocker.

**Responsible disclosure**. We made several attempts to responsibly disclose our findings to the developers of these text messaging apps. Among the 7 apps, VeroSMS responded and acknowledged our findings. Also, VeroSMS admitted that *their poor detection performance is due to the limitation of their training data and the challenge in collecting up-to-date spam messages from their clients due to the privacy constraints enforced by iOS [17]*. Such privacy policies prohibit their app from uploading users' SMS messages to their server, so they cannot retrain their model on the data from different users.

## 6 DISCUSSION

**Limitations of our dataset.** The SMS spam dataset generated by the *SpamHunter* framework contains around 5% of non-spam noise. We believe that such a small ratio should not affect our measurement results. Most such non-spam noise has been introduced by the SMS screenshots containing both spam and non-spam messages, which may mislead our pipeline to consider all of them as spam texts. This problem could be addressed by a more capable spam classifier, which we will study in the future. Another issue is the coverage of *SpamHunter*. Its SRTC component could miss some messages, particularly when Twitter users themselves are not sure whether a message is indeed spam, which renders the sentiment analysis less effective. In our research, we estimated the coverage of our approach on randomly sampled 1000 messages and found that 7.5% of the spam messages were missed by *SpamHunter*.

**Comparison with previous SMS spam datasets.** We compared our dataset with two most up-to-date public SMS spam datasets (Table 10): *SMS Spam Collection* [23], and *FBS (fake-base-station) SMS Spam Dataset* [56]. As presented in the table, our SMS spam data is collected from a new data source – SMS spam reports on Twitter, which has never been studied before. Thus our dataset is unique, includes up-to-date spam samples (e.g., those related to COVID-19) around the world, and differs from other datasets both in content features that render our samples harder to capture (Section 5.4) and in the message distribution over spam categories (Section 3.1). So far our dataset covers 14,785 unique Twitter users and contains 21,918 messages in 75 languages over the past 4 years, which enables a longitudinal study on SMS spam across languages. By comparison, the spam messages from *SMS Spam Collection* and *FBS Spam SMS Dataset* are either in English or Chinese. Also our dataset is featured by a significantly higher ratio (62%) of fraudulent messages compared with other spam datasets, and thus is more helpful for defending against SMS fraud. Last but not least, our dataset is *ever-growing*, which is continuously updated by *SpamHunter* for supporting anti-spam research and development [18].

Table 10: Comparison among public SMS spam datasets.

|  | Ours | SMS Spam Collection | FBS |
| --- | --- | --- | --- |
| # Spam messages | 21,918 | 747 | 14K |
| Period | 4 years | before 2010 | 3 months |
| Language | Multi-language | English | Chinese |
| Source | Twitter report | UK forum | security app |
| SMS spam type | General | General | Fake base-station |
| SMS fraud ratio | 62% | 32% | 38.2% |
| Extendable | Yes | No | No |

**Defending against poisoning attacks.** *SpamHunter* is meant to collect SMS spam samples from the Twitter users who report the spam messages they receive in good faith. A concern is the risk of data contamination, in which malicious users could post fake SMS spam messages to mislead the spam detector trained on such data. Although the problem is general, it certainly needs serious attention. One possible solution is to utilize a Twitter user's reputation when deciding whether to trust her report. For this purpose, we could profile Twitter users based upon their account information, such as account lifetime, number of followers, etc., which has also been used by the prior research on fake account detection [51]. Another strategy is to check the interactions between the spam-reporting user and the relevant parties such as victim services, law enforcement, cellular carriers, etc., particular these parties' responses, to determine the authenticity of the report. The effectiveness of these approaches and other solutions should be studied down the road.

**Recommendations for mitigating SMS spam.** Our measurement on spam network infrastructures implies that small web hosting providers (e.g., Namecheap) should enforce a more strict inspection on spam hosting activities. And future spam detection should consider the new trends of spam infrastructures such as port-forwarding services and dynamic DNS services. Also, anti-spam effort today should make good use of spam intelligence as reported on Twitter, e.g., law enforcement and cellular carriers may leverage the spam-reporting tweets to timely defend against cross-region SMS spam campaigns. Lastly, the discovered spam phone numbers can contribute the existing phone number blacklists and the spam dataset collected by *SpamHunter* can also be used to fine-tune existing machine-learning-based spam detectors.

## 7 RELATED WORK

**Understanding and detecting SMS spam.** A long line of studies [24, 28, 34, 37, 38, 45, 55] have studied how to detect spam SMS leveraging various methodologies especially SVM and Bayesian network, as summarized in [23, 30, 47]. Several SMS spam/ham datasets [22, 30] have also been released to facilitate future spam detection research. However, as discussed above, these datasets are of a small size and tend to become out-of-dated. Another set of studies [22, 43, 49] focus on profiling SMS spams from various aspects such as categories, the underlying spam campaigns, and the spam network infrastructures. Bradley Reaves et. al [43] first introduced a novel channel, i.e. the public SMS gateways, to understand SMS activities. And recently, Yiming Zhang et. al. [56] researched the spam ecosystem and campaigns on fake-base-station SMS spam in China. In our research, we proposed a novel framework to collect SMS spams from a new source - Twitter SMS spam reports, which was never studied before and is different from any existing SMS spam dataset. We also distilled new findings of up-to-date SMS spam in terms of spam strategies, infrastructures and campaigns.

**Evaluation of SMS spam detection.** Akshay Narayan et. al. [42] evaluated spam countermeasures of popular Android text messaging apps leveraging SMS Spam Collection [23]. However, most apps evaluated under their study were found to be out of operation by June 2021. Therefore, when evaluating text messaging apps (§5.4), we composed an up-to-date new set covering popular apps on both Android and iOS platforms. Also, our evaluation is not limited to text messaging apps, but also covers other two kinds of essential players in the SMS ecosystem, namely bulk SMS services and anti-spam services. Last of all, our infiltration utilized up-to-date SMS spam messages as collected by our *SpamHunter*, which allows our infiltration to have distilled a new finding that text messaging apps can miss many up-to-date SMS spam messages while blocking most historical ones.

**Non-SMS Spam.** In addition to SMS spam, spam activities through other communication channels including Email [25, 27, 29, 57], social networks [31, 36, 50, 51], and instant messaging apps [35], are also investigated. These works are dedicated to spam detection and campaign clustering on the specified communication channel. Among them, Payas Gupta et.al. [36] analyzed the spamming campaigns on Twitter with a focus on phone numbers. Our work is different in that we use user reports on Twitter as a data source to analyze the SMS ecosystem and measure different SMS spam strategies in terms of SMS content and spam infrastructure.

## 8 CONCLUSION

In this paper, we present *SpamHunter*, a pipeline to discover up-to-date SMS spam messages reported on Twitter. Using *SpamHunter*, we are able to collect the largest public SMS spam dataset, including 21,918 messages spanning over four years. The dataset is continuous growing and provides us new insight into ongoing SMS spam activities. Our measurement study on the dataset has brought to light the ever-evolving strategies of SMS spamming, its infrastructure, campaigns and others. Our analysis of critical players in the SMS ecosystem highlights the limitations of today's anti-spam effort and potential future directions.

## 9 ACKNOWLEDGEMENT

We would like to thank the anonymous reviewers for their insightful comments. We also appreciate Yingkun Wang's help when building up the evaluation framework.

# A    SPAM CATEGORIES AND EXAMPLES

Table 12 presents the 12 spam categories we have identified during manual analysis, along with each a typical example.

**Define SMS spam categories**. We first define two main SMS spam categories, *Fraud* and *Ads*. A *fraudulent* message tries to deceive SMS recipients to take actions such as clicking some URL or calling a number, and an *Ads* message aims to promote goods/service or propagate opinions. Under these two categories, we further define

Table 11: Reply rate of top tagged Twitter accounts

| Entity | Twitter account | Type | # tagged tweets | Reply rate |
| --- | --- | --- | --- | --- |
| Telkomsel | @Telkomsel | Cellular carrier | 901 | 0.72 |
| ZestMoney | @ZestMoney | Other | 542 | 0 |
| ZestMoney | @ZestMoneyCares | Other | 428 | 0.08 |
| Telecom Regulatory Authority of India | @TRAI | Law-enforcement | 416 | 0 |
| Paytm | @Paytm | Victim service (digital payment) | 414 | 0 |
| Rabobank | @Rabobank | Victim service (bank) | 376 | 0.86 |
| Kominfo | @kemkominfo | Law-enforcement | 355 | 0.01 |
| Policía Nacional España | @polica | Law-enforcement | 340 | 0 |
| State Bank of India | @TheOfficialSBI | Victim service (bank) | 339 | 0.84 |
| IM3 Ooredoo | @IndosatCare | Cellular carrier | 277 | 0.60 |
| Cyber Crime | @Cyberdost | Law-enforcement | 221 | 0.01 |
| Airtel | @Airtel_Presence | Cellular carrier | 221 | 0.79 |

12 subcategories based upon the detailed topic of each message's content. Specifically, in terms of SMS fraud, *Account alert (Fraud)* messages deceive users with a falsified account warning such as suspension or unusual activity; *Finance (Fraud)* messages utilize fake financial transactions or payments to defraud users; *Prize (Fraud)* messages play a traditional trick of winning a prize congratulations; *Delivery (Fraud)* messages pretend to be a delivery notice; *Credit/Debit card (Fraud)* messages pretend a credit/debit card renew/blocking message; *Tax refund (Fraud)* and *COVID-19 (Fraud)* messages utilize recent incidents such as tax refund or COVID-19 pandemic to deceive users; *Other (Fraud)* messages refers to *fraudulent* messages that belong to none of the above subcategories. As for *Ads* messages, *Promotion (Ads)* refers to those marketing messages for promoting goods/services; *Loan/Gamble (Ads)* specifies messages introducing loan or gamble services; *Politics (Ads)* messages are used to propagate political opinions or voting candidates; and *Other (Ads)* messages belong to none of the subcategories in *Ads*.

## B TOP URL SHORTENING SERVICES

Table 14 lists the top 20 URL shorteners that were most adopted by spam campaigns. Table 15 lists the distribution of spam IPs. Table 16 lists the top10 hosting providers of identified spam IPs.

## C SPAM CAMPAIGNS

Figure 6 shows a spam campaign across 4 natural languages.

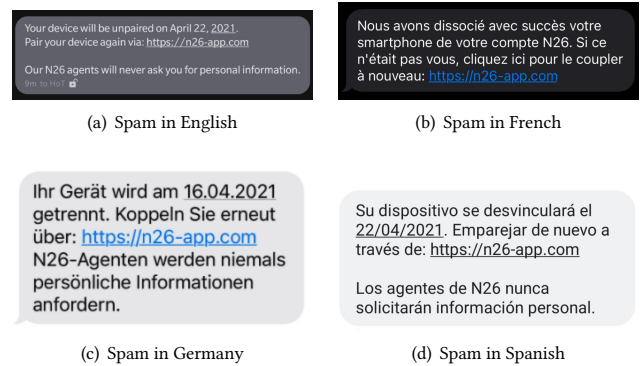

(a) Spam in English    (b) Spam in French

(c) Spam in Germany    (d) Spam in Spanish

Figure 6: A spam campaign across languages.

Table 12: Labeled SMS spam messages dataset

| Category | # Labelled messages | Example |
|---|---|---|
| Account alert (Fraud) | 233 | Hello Customer, Your PAYTM Account is expired & Blocked Today. Please Complete Full Verification Call Customer Care: 620****654 Regard, PAYTM Team |
| Finance (Fraud) | 96 | Triodos Bank: Your authorization for EUR 1601.48 to VOF Spilda has been approved. If you are not familiar with this direct debit, cancel it now via: rplg.co/TRIODOS |
| Prize (Fraud) | 78 | We are trying to get in touch with you - You got 3rd place in our gift on 02/19/2020. Your prize: http://9ig.us/IMYV5 |
| Delivery (Fraud) | 58 | [PostNL]: Your package has been sent. Follow your order via this link: https://sorteercentrum.net/postnl |
| Credit/Debit card (Fraud) | 52 | Rabobank: Your current debit card will expire before use. Request the renewed debit card for free now and prevent blockage via: https://bit.do/Rabobank-NL42 |
| Tax refund (Fraud) | 24 | You have a pending Tax Refund from your council. To proceed with your application please complete your form via www.citycounciilii.com |
| COVID-19 (Fraud) | 16 | Important Coronavirus update in your area http://bitlynow.com/MTM1MTAyNg |
| Other (Fraud) | 33 | It is very necessary that you reach me immediately via edchda45358@gmail.com for details. |
| Promotion (Ads) | 187 | Last 2 Hrs Left to Activate your 85 % Amazon Coupon for your Book Writing & Publishing. Click Here book-writers.com to Activate Now. |
| Loan/Gamble (Ads) | 89 | Sorry to disturb. We from KOPERASI offer loans for those of you who need business capital to serve loans of Rp. 5 million - 300 million U/Info 082196387716 |
| Politics (Ads) | 24 | They'll attack your homes if Joe's elected. Pres Trump needs you to become a Diamond Club Member. Your name is MISSING. Donate: bit.ly/3ipuQPr |
| Other (Ads) | 57 | Wahyuni, I saw you crossing the road earlier, what are you doing in this photo? chat-v.com/s/9tjprx |

Table 15: The distribution of spam IPs.

| Category | # IPs | # Countries | # ASes | # /16 IPv4 | # /8 IPv4 |
|---|---|---|---|---|---|
| All IPs | 33,495 | 93 | 1,353 | 4,060 | 207 |
| $VT \geq 1$ | 5,751 | 65 | 465 | 1,474 | 187 |
| $VT \geq 5$ | 1,142 | 45 | 192 | 498 | 131 |

Table 16: Top hosting providers of spam IPs.

| All IPs Operator | % IP | $VT \geq 1$ Operator | % IP | $VT \geq 5$ Operator | % IP |
|---|---|---|---|---|---|
| Amazon | 39% | Amazon | 28% | Namecheap | 20% |
| Cloudflare | 8% | Cloudflare | 11% | Amazon | 20% |
| Google | 6% | Google | 10% | Cloudflare | 12% |
| Akamai Tech | 4% | Namecheap | 7% | Alibaba | 6% |
| Akamai Intl | 2% | Facebook | 2% | GoDaddy | 2% |
| Namecheap | 2% | Alibaba | 2% | Google | 2% |
| Facebook | 1% | GoDaddy | 2% | DigitalOcean | 1% |
| GoDaddy | 1% | Apple | 2% | Unified Layer | 1% |
| Alibaba | 1% | SFR SA | 2% | Shinjiru Tech | 1% |
| DigitalOcean | 1% | DigitalOcean | 2% | Hostinger | 1% |

Table 14: Top 20 URL shorteners used most frequently in SMS spam campaigns.

| Shortener Name | # Spam URLs | # SMS Spam Messages |
|---|---|---|
| bit.ly | 2,390 | 2,889 |
| s.id | 223 | 271 |
| bit.do | 149 | 184 |
| wa.me | 145 | 150 |
| cutt.ly | 121 | 147 |
| tsel.me | 119 | 127 |
| chat-v.com | 99 | 99 |
| goo.gl | 96 | 100 |
| www.bit.ly | 84 | 96 |
| tiny.cc | 77 | 117 |
| tinyurl.com | 72 | 83 |
| is.gd | 63 | 784 |
| nmc.sg | 45 | 41 |
| x.co | 43 | 81 |
| linksplit.io | 36 | 55 |
| t.co | 31 | 35 |
| lihi1.cc | 30 | 33 |
| tny.sh | 28 | 30 |
| rebrand.ly | 25 | 25 |
| rb.gy | 23 | 30 |

Table 13: The statistics of time gap in days between a spam reporting on Twitter and its VirusTotal reports. $T_{gap} = Date_{VT} - Date_{Twitter}$, VT-M is the VT category of malicious, VT-MW is the VT category of malware and VT-P is the VT category of phishing.

| URL Group | $\geq -7$ | $\geq -1$ | $\geq 0$ | $\geq 1$ | $\geq 7$ |
|---|---|---|---|---|---|
| VT >= 1 | 83.5% | 75.8% | 66.9% | 30.1% | 15.4% |
| VT >= 5 | 89.5% | 81.8% | 70.7% | 27.7% | 10.1% |
| VT >= 10 | 91.2% | 84.7% | 74.0% | 22.8% | 5.1% |
| VT-M | 84.9% | 77.8% | 68.1% | 29.7% | 14.9% |
| VT-MW | 84.9% | 76.2% | 67.0% | 24.1% | 9.4% |
| VT-P | 86.8% | 79.1% | 69.5% | 30.2% | 14.0% |